\documentclass[12pt]{article}

\usepackage{epsf}
\usepackage{epsfig}
\usepackage{cite}
\usepackage{amsmath}
\usepackage{graphics}

\begin{document}

\setlength{\textheight}{21.5cm}
\setlength{\oddsidemargin}{0.cm}
\setlength{\evensidemargin}{0.cm}
\setlength{\topmargin}{0.cm}
\setlength{\footskip}{1cm}
\setlength{\arraycolsep}{2pt}

\renewcommand{\thefootnote}{\#\arabic{footnote}}
\setcounter{footnote}{0}

\newcommand{\gtrsim}{ \mathop{}_{\textstyle \sim}^{\textstyle >} }
\newcommand{\lesssim}{ \mathop{}_{\textstyle \sim}^{\textstyle <} }
\newcommand{\rem}[1]{{\bf #1}}
\renewcommand{\thefootnote}{\fnsymbol{footnote}}
\setcounter{footnote}{0}
\def\thefootnote{\fnsymbol{footnote}}

\hfill December 2009\\

\hfill IPMU-09-0123\\

\vskip .5in

\begin{center}

\bigskip
\bigskip

{\Large \bf Alternative Version of Chiral Color as Alternative to the
Standard Model} 

\vskip .45in

{\bf Paul H. Frampton\footnote{frampton@physics.unc.edu}} 

\vskip .3in

{\it Department of Physics and Astronomy, University of North Carolina,
Chapel Hill, NC 27599-3255.\footnote{permanent address}}

{\it Institute for the Physics and Mathematics of the
Universe, University of Tokyo, Kashiwa, Chiba 277-8568, Japan.}

\end{center}

\vskip .4in 
\begin{abstract}
In a variant of chiral color with the 
electroweak gauge group
generalized to $SU(3)_L \times U(1)$ anomaly
cancellation occurs more readily than in
the $SU(2)_L \times U(1)$ case. Three families
are required by anomaly cancellation
and
the top family appears non-sequentially.
\end{abstract}

\renewcommand{\thepage}{\arabic{page}}
\setcounter{page}{1}
\renewcommand{\thefootnote}{\#\arabic{footnote}}

\newpage

\noindent {\it Introduction}

\bigskip

\noindent Any small discrepancies from the standard model of particle
theory merit further study to point the way towards
an extension of the model. One such possible discrepancy
is the motivation for the present Letter.

\bigskip

\noindent Recent experimental data on top production at FermiLab\cite{ttexpt}
give hints about a departure from the predictions of QCD. One 
possible interpretation,
{\it e.g.} \cite{ttpheno},
is in terms of an axigluon in the s-channel of the quark-antiquark
annihilation. Therefore it is worth re-examining that theory from the
1980's with a view to enunciating any additional predictions for
experiment.

\bigskip

\noindent Chiral color represents one path of model
building beyond the standard model \cite{FG}
\footnote{Precursors appeared in \cite{PS}.}.
The color symmetry of QCD results from symmetry
breaking of a $SU(3)_L \times SU(3)_R$ gauge
symmetry at TeV scales and leads to the prediction
of a color octet of massive axigluons. 

\bigskip

\noindent Searches in hadron colliders have led to
a lower limit on the axigluon mass of at least 1 TeV
\cite{UA1CDF}.

\bigskip

\noindent More recently, a small ($2 \sigma$) effect
in the $t\bar{t}$ forward-backward asymmetry
at the Tevatron\cite{ttexpt} has prompted some analysis in terms
of axigluon, as well as other possible new colored
particles\cite{ttpheno}. While this experimental anomaly
has too large a statistical error to be taken
as firmly based, it does lead to a possible re-examination
of the chiral color model.

\bigskip

\noindent Here we show that, by generalizing the original
version\cite{FG}, one can arrive at a simplified anomaly
cancellation which requires the existence of three families.

\bigskip
\bigskip

\noindent {\it Anomalies of Original Chiral Color}.

\bigskip

\noindent First recall that the
original chiral color used the gauge group

\begin{equation}
SU(3)_{CL} \times SU(3)_{CR} \times SU(2)_L \times U(1)_Y
\end{equation}

\noindent with each family assigned to the reducible representation

\bigskip

(3, 1; 2, 1/3) + (1, 3; 2, 1/3)

+ (3*, 1; 1, -4/3) + (3*, 1; 1, 2/3)

+ (1, 3*;1, 2/3) + (1, 3*; -4/3)

+ (1, 1; 2, -1)

+ (1, 1; 1, +2)

\bigskip

\noindent This model needed additional fermions to cancel anomalies because for
the seven potential triangle anomalies for each sequential family:

(I) $Y^3$  fails.

(II) $Y 3_{CL}^2$  cancels. 

(III) $Y 3_{CR}^2$ cancels. 

(IV) $Y 2_L^2$  fails. 

(V) $3_{CL}^3$ cancels.

(VI) $3_{CR}^3$ cancels. 

(VII) Y  cancels.

\bigskip

\noindent Resolving the anomaly cancellation for
(I) and (IV) led to a variety or proposals for
adding further colored chiral fermion\cite{FG}

\bigskip
\bigskip

\noindent {\it Anomalies of Alternative Chiral Color}.

\bigskip

\noindent The gauge group in the generalization will be
\footnote{This is related to models in \cite{SU15}.}
\begin{equation}
SU(3)_{CL} \times SU(3)_{CR} \times SU(3)_L \times U(1)_X
\end{equation}

\noindent The 1st family is assigned as follows

\bigskip

LH quarks    (3, 1; 3, -1/3)

RH quarks  (1, 3*; 1, -2/3) + (1,3*; 1, 1/3) +  1/3) + (1, 3*;1, 4/3)

leptons  (1, 1; 3*, 0)

\bigskip

\noindent and the second family is

\bigskip

LH quarks    (1, 3; 3, -1/3)

RH quarks  (3*, 1; 1, -2/3) + (3*, 1; 1, 1/3) +  1/3) + (3*, 1 ;1, 4/3)

leptons  (1, 1; 3*, 0)

\bigskip

\noindent while the third family is

\bigskip

LH quarks    (3, 1; 3*, 2/3) 

RH quarks  (3*, 1; 1, -5/3) + (3*, 1; 1, -2/3) + (3*, 1; 1, 1/3)

leptons  (1, 1; 3*, 0)

\bigskip

\noindent This introduces an anomaly (VIII)  $(3_L)^3$
beyond the analogs of (I) through (VII)
for the original chiral color model
where there is no $(2_L)^3$ anomaly. 
With the assigmements given above {\it all}
anomalies (I) through (VIII) cancel.

\bigskip

\noindent What is remarkable is that
this alternative chiral color
succeeds to cancel the analogs of \underline{all} the seven
anomalies (I) - (VII) above, plus the
new anomaly (VIII) $3_L^3$; also {\it e.g.} the $X^3$ anomaly
requires three families for cancellation.

\bigskip

\noindent {\it Higgs sector}

\bigskip

\noindent In order to break chiral color to QCD
we require a VEV of a bi-triplet scalar

\bigskip

\noindent (3, 3*; 1, 0)

\bigskip

\noindent while to break the electroweak group
$SU(3)_L \times U(1)_X \rightarrow SU(2)_L \times U(1)_Y
\rightarrow U(1)_{EM}$ requires the scalars

\bigskip

\noindent (1, 1; 3, 0) + (1, 1; 3, 1)

\bigskip

\noindent whose VEVs can provide mass to all
the quarks and charged leptons.

\bigskip
\bigskip
\bigskip

\noindent {\it Discussion}

\bigskip

\noindent Partially because of the hints from $p\bar{p}
\rightarrow t\bar{t}$ (and they are no more than hints)
that new colored objects may be necessary, it seems worth
re-examining models which have this feature.

\bigskip

\noindent The present alternative to chiral color
provides a novel three-family anomaly cancellation
and places the third family containing the $t$-quark
aymmetrically with respect to lighter quarks.

\bigskip

\noindent The present alternative version predicts additionally
the bilepton gauge bosons discussed in \cite{SU15}
necessary for the simpler arrangement of anomaly
cancellation, so that there are gauge bosons
beyond those of the standard model in both
the strong and electroweak sectors separately.

\bigskip

\noindent It will be interesting to see whether the
effects reported in \cite{ttexpt} will survive
as the experiments acquire
higher statistics. Beyond that, there
is the question whether
additional gauge bosons as predicted here in 
the electroweak sector will be discovered.  

\newpage

\begin{center}

\section*{Acknowledgements}

\end{center}

This work was supported in part 
by the U.S. Department of Energy under Grant
No. DE-FG02-06ER41418.


\newpage

\bigskip
\bigskip
\bigskip


\begin{thebibliography}{99}
\bibitem{ttexpt}
T. Aaltonen, {\it at al},(CDF Collaboration),
Phys. Rev. lett. {\bf 101,} 202001 (2008).
\\
V. M. Abazov {\it et al}, (D0 Collaboration), Phys. Rev. Lett. {\bf 100,}
142002(2008);\\
http://www-cdf.fnal.gov/physics/new/top/2009/tprop/Afb/

\bibitem{ttpheno}
P. Ferrario and G. Rodrigo, Phys. Rev.{\bf D78,} 094018 (2008)
{\tt arXiv: 0809.3354 [hep-ph]};\\
Phys. Rev. {\bf D80,} 051701 (2009). {\tt arXiv:0906.5541 [hep-ph]}.\\
P.H. Frampton, J. Shu and K. Wang, Phys. Lett. {\bf B} (in press)
{\tt arXiv:0911.2955 [hep-ph]}.
\bibitem{FG}
P.H. Frampton and S.L. Glashow, Phys.Lett. {\bf 190B,} 157 (1987);\\
Phys. Rev. Lett. {\bf 58,} 2168 (1987).
\bibitem{PS}
J. C. Pati and A. Salam, Phys. Lett. {\bf B58,} 333 (1975);\\
L.J. Hall and A.E. Nelson, Phys. Lett. {\bf B153,} 430 (1985).
\bibitem{UA1CDF}
C. Albajar et al. (UA1 Collaboration), Phys. Lett. B209, 127 (1988);\\
F. Abe et al. (CDF Collaboration) Phys. Rev. D41, 1722 (1990);\\
Phys.Rev. D55, R5263 (1997).
\bibitem{SU15}
P.H. Frampton and B.H. Lee, Phys. Rev. Lett. {\bf 64,} 619 (1990);\\
P.H. Frampton, Phys. Rev.Lett. {\bf 69,} 889 (1992).\\
F. Pisano and V. Pleitez, Phys. Rev. {\bf D46,} 410 (1992).

\end{thebibliography}
\end{document}